\documentstyle[12pt]{article}
\newcommand{\newsection}{    
\setcounter{equation}{0}
\section}
\def\appendix#1{
  \addtocounter{section}{1}
  \setcounter{equation}{0}
  \renewcommand{\thesection}{\Alph{section}}
  \section*{Appendix \thesection\protect\indent #1}
  \addcontentsline{toc}{section}{Appendix \thesection\ \ \ #1}
  }
\newcommand{\tr}[1]{\,{\rm tr}#1}

\def\e{{\,\rm e}\,}

\def\eop{\vspace*{\fill}\pagebreak}
\def\be{\begin{equation}}
\def\ee{\end{equation}}
\def\bea{\begin{eqnarray}}
\def\eea{\end{eqnarray}}
\def\LA{\left\langle}
\def\RA{\right\rangle}

\newcommand{\rf}[1]{(\ref{#1})}
\newcommand{\eq}[1]{Eq.~(\ref{#1})}

\def\l{\lambda}

\def\om{\omega}

\newcommand{\p}{{\prime}}
\newcommand{\ra}{\rightarrow}
\newcommand{\fr}[2]{{\textstyle {#1 \over #2}}}

\newcommand{\non}{\nonumber \\*}

\newcommand{\im}{\,\hbox{Im}\,}

\def\fun#1#2{\lower3.6pt\vbox{\baselineskip0pt\lineskip.9pt
\ialign{$\mathsurround=0pt#1\hfil##\hfil$\crcr#2\crcr\sim\crcr}}}

\begin{document}

\begin{titlepage}
\begin{flushright}
NBI-HE-94-25 \\
hep-ph/9404312 \\
April, 1994
\end{flushright}
\vspace{.5cm}

\begin{center}
{\LARGE Exact Multiparticle Amplitudes at Threshold} \\
 \vspace{0.6cm}{\LARGE in Large-$N$ Component $\phi^4$ Theory}
\end{center} \vspace{1cm}
\begin{center}
{\large Yu.\ Makeenko}\footnote{E--mail: \ makeenko@nbivax.nbi.dk \ / \
 makeenko@vxitep.itep.msk.su \ }
\\ \mbox{} \\
{\it The Niels Bohr Institute,} \\
{\it Blegdamsvej 17, 2100 Copenhagen, DK} \\ \vskip .2 cm
and  \\  \vskip .2 cm
{\it Institute of Theoretical and Experimental Physics,}
\\ {\it B. Cheremushkinskaya 25, 117259 Moscow, RF}
\end{center}

\vskip 1 cm
\begin{abstract}
I derive the set of recurrence relations between the amplitudes of
multiparticle production at threshold
in the standard large-$N$ limit of the $O(N)$-symmetric $\phi^4$ theory
which sums all relevant diagrams with arbitrary number of loops.
I find an exact solution to the recurrence relations using the
Gelfand--Diki\u{\i} representation of the diagonal resolvent
of the Schr\"{o}dinger operator. The result coincides with the
tree amplitudes while the effect of loops is the renormalization of
the coupling constant and mass. The form of the solution
is due to the fact that the exact amplitude of the process $2$$\ra$$n$
vanishes at $n$$>$$2$ on mass shell
when averaged over the $O(N)$-indices of incoming particles
for dynamical reasons because of the cancellation between diagrams.
I discuss some possible applications of large-$N$ amplitudes,
in particular, for the renormalon problem.
\end{abstract}

\vspace{.5cm}
\noindent
Submitted to {\sl Physical Review D}

\eop
\end{titlepage}
\setcounter{page}{2}

\newsection{Introduction}

The problem of calculating amplitudes of multiparticle production
at threshold has recently received a considerable
interest~\cite{Vol92}--\cite{LRT93b}. The explicit results for the
tree level amplitudes in $\phi^4$ theory demonstrate the
factorial grows which is expected due to the large amount
of identical bosons in the final state. Further investigations of
the tree amplitudes~\cite{AKP93a,Bro92} introduced
a nice technique to deal with the problem.
A very interesting property of the tree amplitudes which has been
recently pointed out by Voloshin~\cite{Vol93a} is the nullification
of the on-mass-shell amplitude of the process $2$$\ra$$n$ for
$n$$>$$4$. This nullification has been
extended~\cite{Smi93}--\cite{LRT93a} to
more general models and made it possible to calculate the
amplitude $1$$\ra$$n$ at the one-loop level.
A dynamical symmetry which may be responsible for the nullification
has been discussed by Libanov~et~al.~\cite{LRT93b}.

In the present paper I shall make an attempt to
understand which properties of the tree and one-loop
amplitudes could survive in the full theory which
includes all loop diagrams. For this purpose I consider
the $N$-component $\phi^4$ theory which is exactly
solvable in the large-$N$ limit~\cite{Wil73} when only the bubble
diagrams contribute if there is no multiparticle production.

I derive the set of recurrence relations between the amplitudes of
multiparticle production at threshold
in the large-$N$ limit of the $O(N)$-symmetric $\phi^4$ theory
which sums all relevant diagrams with an arbitrary number of loops.
I reduce these recurrence relations to a quantum mechanical
problem which turns out to be
 possible due to the factorization at large $N$.
I find an exact solution to the problem using the
Gelfand--Diki\u{\i} representation of the diagonal resolvent
of the Schr\"{o}dinger operator. The result  is quite similar to the
tree amplitudes while the effect of loops is the renormalization of
the coupling constant and mass. The form of the solution
is due to the fact that the exact amplitude of the process $2$$\ra$$n$
vanishes for $n$$>$$2$ at large $N$
on mass shell when averaged over the $O(N)$-indices of
incoming particles. This nullification occurs
for dynamical reasons because of the cancellation between diagrams.

This paper is organized as follows. Sect.~2 is devoted to the definitions
and the description of the kinematics.
In Sect.~3 I derive the set of the recurrence relations and
transform it to a quantum mechanical problem.
In Sect.~4 I present an exact solution and demonstrate that it
satisfies the set of equations.  The exact solution is possible
since the diagonal resolvent of the Schr\"{o}dinger operator
has a very simple form. I extend the method of
calculating the diagonal resolvent to the case of a more general
potential of the P\"{o}schl--Teller type.
In Sect.~5 I consider some implications of the results
and discuss why the calculation of large-$N$ amplitudes for
multiparticle production at threshold can be interesting
for the renormalon problem.
Appendix~A contains a proof of the uniqueness of the solution.

\newsection{The definitions}

The $O(N)$-symmetric $\phi^4$ theory is defined in the
$4$-dimensional Minkowski space
by the following Lagrangian
\be
{\cal L} = \fr 12 (\partial_\mu \phi^b) (\partial_\mu \phi^b)
-\fr 12 m^2({\phi}^b{\phi}^b)
-\fr 14 \l ({\phi}^b{\phi}^b)^2
\label{lagrangian}
\ee
where $b=1,\ldots,N$ and the summation over repeated indices is implied.

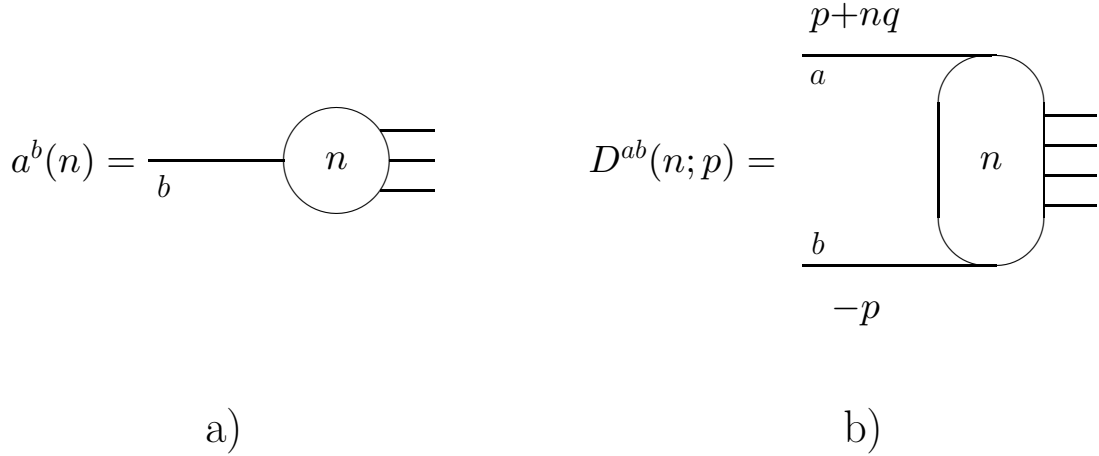
\begin{figure}[tbp]
\unitlength=1.00mm
\linethickness{0.6pt}
\centering
\begin{picture}(118.00,68.00)(0,70)
\put(15.00,78.00){\makebox(0,0)[cc]{{\Large a)}}}
\put(30.00,114.00){\circle{14.00}}
\put(5.00,114.00){\line(1,0){18.00}}
\put(7.00,112.00){\makebox(0,0)[ct]{$b$}}
\put(37.00,114.00){\line(1,0){6.00}}
\put(35.80,118.00){\line(1,0){7.20}}
\put(35.80,110.00){\line(1,0){7.20}}
\put(30.00,114.00){\makebox(0,0)[cc]{{\large $n$}}}
\put(-5.00,114.00){\makebox(0,0)[cc]{{\large $a^b(n)=$}}}
\put(100.00,78.00){\makebox(0,0)[cc]{{\Large b)}}}
\put(117.00,114.00){\oval(14.00,28.00)[]}
\put(117.00,128.00){\line(-1,0){25.00}}
\put(117.00,100.00){\line(-1,0){25.00}}
\put(117.00,114.00){\makebox(0,0)[cc]{{\large $n$}}}
\put(124.00,116.00){\line(1,0){7.00}}
\put(124.00,112.00){\line(1,0){7.00}}
\put(124.00,108.00){\line(1,0){7.00}}
\put(124.00,120.00){\line(1,0){7.00}}
\put(99.00,132.00){\makebox(0,0)[cb]{{\large $p$$+$$nq$}}}
\put(99.00,96.00){\makebox(0,0)[ct]{{\large $-$$p$}}}
\put(76.00,114.00){\makebox(0,0)[cc]{{\large $D^{ab}(n;p) =$}}}
\put(94.00,126.00){\makebox(0,0)[ct]{$a$}}
\put(94.00,103.00){\makebox(0,0)[cc]{$b$}}
\end{picture}
\caption[x]   {\hspace{0.2cm}\parbox[t]{13cm}
{\small
   The graphic representations of the multiparticle
   amplitudes $a^b(n)$ (\/{\normalsize a)}\/) and $D^{ab}(n;p)$
   (\/{\normalsize b)}\/). }}
\label{fig1}
\end{figure}

We denote by $a^b_{b_1\cdots b_n}(n)$ the amplitude of production of $n$
on-mass-shell particles with the $O(N)$-indices
${b_1\ldots b_n}$ at rest by a (virtual) particle
with the $O(N)$-index $b$ and the energy $nm$:
\be
a^b_{b_1\cdots b_n}(n ) = \LA b_1\ldots b_n| \phi^b(0) |0 \RA \,.
\ee
It is convenient to multiply this amplitude by
the vectors $\xi^{b_i}$ which results in a symmetrization over the
$O(N)$-indices of the produced particles and define
\be
a^b(n) = a^b_{b_1\cdots b_n}(n)\, \xi^{b_1}\ldots \xi^{b_n} \,.
\label{defa}
\ee
The amplitude $a^b(n)$ is depicted graphically in Fig.~\ref{fig1}$a$.

The second quantity which appears in the set of the recurrence
relations between the amplitudes is the amplitude
$D^{ab}_{b_1\cdots b_n}(n;p)$ of the process
when two particles $a$ and $b$ with the $4$-momenta
$p+nq$ and $-p$, respectively, produce $n$ on-mass-shell particles
with the $O(N)$-indices ${b_1\ldots b_n}$
and the equal $4$-momenta $q=(m,0)$
in the rest frame. We define again
\be
D^{ab}(n;p)=D^{ab}_{b_1\cdots b_n}(n;p)\, \xi^{b_1}\ldots \xi^{b_n} \,.
\label{defD}
\ee
The amplitude $D^{ab}(n;p)$ is depicted graphically in
Fig.~\ref{fig1}$b$.
The definitions~\rf{defa} and~\rf{defD} coincide for $N=1$ with those
originally introduced by Voloshin~\cite{Vol92,Vol93a}.

It is easy to estimate at the tree level that
\be
a^b(n) \sim \left(\l{\xi}^2\right)^{\frac{n-1}{2}} \xi^b
\label{order}
\ee
where $\xi^2$ stands for $\xi^a\xi^a$. Since
\be
\l \sim \frac 1N
\label{orderl}
\ee
in the large-$N$ limit~\cite{Wil73}, we choose
\be
{\xi}^2 \sim N
\label{orderxi}
\ee
for all the amplitudes~\rf{order} to be of the same order in $1/N$.

\newsection{Recurrence relations at large $N$}

The condition~\rf{orderl} leaves at large-$N$ only the diagrams with
the largest possible number of sums over internal $O(N)$-indices which
propagate along closed loops of diagrams.
The proper recurrence relations
which extend the usual Schwinger--Dyson equations to the case when
$n$ particles are produced are
depicted for $a^b(n)$ in Fig.~\ref{fig3}.
\begin{figure}[tbp]
\unitlength=1.00mm
\linethickness{0.6pt}
\centering
\begin{picture}(108.00,58.00)(41,80)
\put(40.00,114.00){\circle{14.00}}
\put(17.00,114.00){\line(1,0){16.00}}
\put(47.00,114.00){\line(1,0){6.00}}
\put(45.80,118.00){\line(1,0){7.20}}
\put(45.80,110.00){\line(1,0){7.20}}
\put(40.00,114.00){\makebox(0,0)[cc]{{\large $n$}}}
\put(61.00,114.00){\makebox(0,0)[cc]{$=$}}
\put(68.00,114.00){\line(1,0){18.00}}
\put(86.00,114.00){\line(1,1){13.90}}
\put(103.00,131.00){\circle{8.00}}
\put(103.00,131.00){\makebox(0,0)[cc]{$n_1$}}
\put(106.00,134.00){\line(1,0){7.00}}
\put(107.15,131.00){\line(1,0){5.85}}
\put(106.00,128.00){\line(1,0){7.00}}
\put(87.00,113.00){\line(1,0){11.90}}
\put(103.00,113.00){\circle{8.00}}
\put(103.00,113.00){\makebox(0,0)[cc]{$n_2$}}
\put(106.00,116.00){\line(1,0){7.00}}
\put(107.15,113.00){\line(1,0){5.85}}
\put(106.00,110.00){\line(1,0){7.00}}
\put(87.00,113.00){\line(1,-1){13.00}}
\put(103.00,97.00){\circle{8.00}}
\put(103.00,97.00){\makebox(0,0)[cc]{$n_3$}}
\put(106.00,100.00){\line(1,0){7.00}}
\put(107.15,97.00){\line(1,0){5.85}}
\put(106.00,94.00){\line(1,0){7.00}}
\put(121.00,114.00){\makebox(0,0)[cc]{$+$}}
\put(128.00,114.00){\line(1,0){18.00}}
\put(146.00,114.00){\line(1,1){13.90}}
\put(163.00,131.00){\circle{8.00}}
\put(163.00,131.00){\makebox(0,0)[cc]{$n_1$}}
\put(166.00,134.00){\line(1,0){7.00}}
\put(167.15,131.00){\line(1,0){5.85}}
\put(166.00,127.83){\line(1,0){7.00}}
\put(163.00,106.00){\oval(8.00,14.00)[]}
\put(163.00,106.00){\makebox(0,0)[cc]{$n_2$}}
\put(147.00,113.00){\line(1,0){16.00}}
\put(147.00,113.00){\line(1,-1){13.00}}
\put(167.00,104.50){\line(1,0){6.00}}
\put(167.00,107.50){\line(1,0){6.00}}
\put(166.85,110.50){\line(1,0){6.15}}
\put(166.85,101.50){\line(1,0){6.15}}
\end{picture}
\caption[x]   {\hspace{0.2cm}\parbox[t]{13cm}
{\small
   The recurrence relation for the amplitude $a^b(n)$.
   The analytical formula is given by \eq{recurrencea}.}}
\label{fig3}
\end{figure}
 The continuous lines are associated
with  propagation of the $O(N)$-indices.
Each vertex is the sum of three possible permutations
of the $O(N)$-indices.  This is taken
into account in Fig.~\ref{fig3} by a combinatorial factor.
The graphic notations becomes clear if one  introduces
the auxiliary field $\sigma(x)=\phi^2(x)$
which propagates in the empty space inside vertices.

The recurrence relation of Fig.~\ref{fig3}
reads analytically as
\bea
a^a(n) = \l m^{-6} \sum_{n_1,n_2,n_3 = odd}
\delta_{n,n_1+n_2+n_3} \frac{n!}{n_1!n_2!n_3!}
\frac{a^a(n_1)a^b(n_2)a^b(n_3)}{(n^2_1-1)(n^2_2-1)(n^2_3-1)} \non
+\l m^{-2} \sum_{n_1=odd,\atop n_2=even}
\delta_{n,n_1+n_2} \frac{n!}{n_1!n_2!}
\frac{a^a(n_1)}{(n^2_1-1)}\int \frac{d^4 k}{(2\pi)^4}
D^{bb}(n_2;k) \,.
\label{recurrencea}
\eea
Notice that only the summed over the $O(N)$-indices quantity
$D^{bb}(n_2;k)$ enters this recurrence relation.

It is the property of large $N$ that the recurrence relation
for $D^{aa}(n;p)$ expresses it via $a^b$ and $D^{bb}$
again. The recurrence relation
for $D^{aa}(n;p)$ is depicted in Fig.~\ref{fig4}.
\begin{figure}[tbp]
\unitlength=1.00mm
\linethickness{0.6pt}
\centering
\begin{picture}(108.00,63.00)(41,80)
\put(40.00,114.00){\oval(14.00,28.00)[]}
\put(40.00,128.00){\line(-1,0){22.00}}
\put(40.00,100.00){\line(-1,0){22.00}}
\put(40.00,114.00){\makebox(0,0)[cc]{{\large $n$}}}
\put(47.00,116.00){\line(1,0){7.00}}
\put(47.00,112.00){\line(1,0){7.00}}
\put(47.00,108.00){\line(1,0){7.00}}
\put(47.00,120.00){\line(1,0){7.00}}
\put(61.00,114.00){\makebox(0,0)[cc]{$=$}}
\put(68.00,128.00){\line(1,0){18.00}}
\put(87.50,128.00){\line(2,1){12.00}}
\put(103.50,136.00){\circle{8.00}}
\put(103.50,136.00){\makebox(0,0)[cc]{$n_1$}}
\put(106.50,139.00){\line(1,0){7.00}}
\put(107.65,136.00){\line(1,0){5.85}}
\put(106.50,133.00){\line(1,0){7.00}}
\put(87.50,128.00){\line(2,-1){12.00}}
\put(103.50,120.00){\circle{8.00}}
\put(103.50,120.00){\makebox(0,0)[cc]{$n_2$}}
\put(106.50,123.00){\line(1,0){7.00}}
\put(107.65,120.00){\line(1,0){5.85}}
\put(106.50,117.00){\line(1,0){7.00}}
\put(68.00,100.00){\line(1,0){18.00}}
\put(86.00,107.00){\oval(8.00,14.00)[]}
\put(86.00,107.00){\makebox(0,0)[cc]{$n_3$}}
\put(86.00,114.00){\line(0,1){14.00}}
\put(90.00,105.50){\line(1,0){6.00}}
\put(90.00,108.50){\line(1,0){6.00}}
\put(89.85,111.50){\line(1,0){6.15}}
\put(89.85,102.50){\line(1,0){6.15}}
\put(121.00,114.00){\makebox(0,0)[cc]{$+$}}
\put(128.00,128.00){\line(1,0){18.00}}
\put(162.50,128.00){\oval(8.00,14.00)[]}
\put(162.50,128.00){\makebox(0,0)[cc]{$n_1$}}
\put(166.50,126.50){\line(1,0){6.00}}
\put(166.50,129.50){\line(1,0){6.00}}
\put(166.35,132.50){\line(1,0){6.15}}
\put(166.35,123.50){\line(1,0){6.15}}
\put(147.50,128.00){\line(2,1){13.00}}
\put(147.50,128.00){\line(2,-1){13.00}}
\put(128.00,100.00){\line(1,0){18.00}}
\put(146.00,107.00){\oval(8.00,14.00)[]}
\put(146.00,107.00){\makebox(0,0)[cc]{$n_2$}}
\put(146.00,114.00){\line(0,1){14.00}}
\put(150.00,105.50){\line(1,0){6.00}}
\put(150.00,108.50){\line(1,0){6.00}}
\put(149.85,111.50){\line(1,0){6.15}}
\put(149.85,102.50){\line(1,0){6.15}}
\end{picture}
\caption[x]   {\hspace{0.2cm}\parbox[t]{13cm}
{\small
   The recurrence relation for the amplitude $D^{aa}(n;p)$.
   The analytical formula is given by \eq{recurrenceD}.
   }}
\label{fig4}
\end{figure}

The recurrence relation of Fig.~\ref{fig4}
reads analytically as%
\footnote{Here and below the poles should be understood
according to the Feynman prescription $m^2\ra m^2-i0$.}
\bea
D^{aa}(n) = \frac{\l m^{-4}}{(p+nq)^2-m^2}
 \sum_{n_1,n_2=odd, \atop n_3 = even}
\delta_{n,n_1+n_2+n_3} \frac{n!}{n_1!n_2!n_3!}
\frac{a^b(n_1)a^b(n_2)D^{aa}(n_3;p)}{(n_1^2-1)(n_2^2-1)}
\non +\frac{\l}{(p+nq)^2-m^2} \sum_{n_1,n_2=even}
\delta_{n,n_1+n_2} \frac{n!}{n_1!n_2!}
\int \frac{d^4 k}{(2\pi)^4} D^{bb}(n_1;k) D^{aa}(n_2;p) \,.
\label{recurrenceD}
\eea

Eqs.~\rf{recurrencea} and \rf{recurrenceD}
look very similar to the ones~\cite{Vol93a,AKP93c} for
$N=1$ at the one-loop level for $a(n)$ when one needs $D(n;p)$
only at the tree level so that the second term on the r.h.s.\ of
\eq{recurrenceD} can be omitted.

To rewrite Eqs.~\rf{recurrencea} and \rf{recurrenceD} in a more
convenient form, let us introduce,
following the approach of Argyres et al.~\cite{AKP93a},
the generating functions
\be
\Phi^a(\tau) = m \xi^a \e^{m\tau}  + \sum_{n\geq3}
 \frac{a^a(n)}{n!(n^2-1)} \e^{nm\tau} m^{n-2}
\label{Phi}
\ee
and
\be
D^{ab}(\tau;p) = \frac{\delta^{ab}}{p^2-m^2}  + \sum_{n=even}
D^{ab}(n;p) \frac{1}{n!} \e^{nm\tau} m^{n} \,.
\label{D}
\ee

Eqs.~\rf{recurrencea} and \rf{recurrenceD} can then be rewritten,
respectively, as
\be
\left\{\frac{d^2}{d\tau^2} -m^2- v(\tau)\right \} \Phi^a(\tau) =0
\label{eq1}
\ee
and
\be
\left\{\frac{d^2}{d\tau^2} -\om^2- v(\tau)  \right\}
\e^{\epsilon \tau}\frac 1N D^{bb}(\tau;p)=
\e^{\epsilon \tau}
\label{eq2}
\ee
where $\epsilon$ is the energy component of $p$,
$p=(\epsilon,\vec{p}\,)$,
\be
\om = \sqrt{\vec{p}\;{}^2+m^2}
\ee
and
\be
v(\tau) = \l \Phi^2(\tau)+\l \int \frac{d^4
k}{(2\pi)^4}D^{bb}(\tau;k) \,.
\label{u}
\ee
This transformation from  Eqs.~\rf{recurrencea} and \rf{recurrenceD} to
Eqs.~\rf{eq1} and \rf{eq2} is
quite similar to the one~\cite{Vol93a,AKP93c}
for $N=1$ at the one-loop level. The
extension to the $O(N)$ case was considered
by Brown~\cite{Bro92} at the tree level and by Smith~\cite{Smi93a}
at the one loop level.

It is easy to understand using \eq{D}
that $v(\tau)$ which is defined by \eq{u} is nothing but the sum of
the matrix elements:
\be
v(\tau) = \l \sum_{n=0}^\infty
\LA  b_1\ldots b_n| \phi^2(0)|0 \RA \xi^{b_1} \ldots \xi^{b_n}
\frac{m^n}{n!}
\e^{nm\tau} \,.
\label{ume}
\ee
This formula as well as Eqs.~\rf{eq1}, \rf{eq2}
can alternatively be derived by the use of the functional
technique introduced for this problem by Brown~\cite{Bro92}
which relates $\tau$ to the time variable $t\equiv x_0$ of $\phi^a(x)$
by
\be
\tau = it \,.
\ee

The fact that the matrix elements rather than operators themselves
appear in Eqs.~\rf{eq1} and \rf{eq2} \sloppy
is due to the factorization at large $N$  (see e.g.\ Ref.~\cite{Mak83}).
It is the reason why Eqs.~\rf{eq1} and \rf{eq2} form a
closed set at large $N$.

The next step in the transformation of
Eqs.~\rf{eq1}, \rf{eq2} and \rf{u} is to pass for $D^{ab}$ to the mixed
representation --- the coordinate in (imaginary) time and
momentum for space. One defines
\be
D^{ab}_\om(\tau,\tau^\prime) = \int \frac{d\epsilon}{2\pi}
\e^{\epsilon(\tau-\tau^\prime)} D^{ab}(\tau;p) \,.
\label{Fourier}
\ee
In order to see that this quantity is indeed associated with the
Fourier transform of  amplitudes in energy, we notice that the
insertion of \eq{D} on the r.h.s.\ of \eq{Fourier} yields
\be
D^{ab}_\om(\tau,\tau^\prime) = \int \frac{d\epsilon}{2\pi}
\sum_{n=0}^\infty
\e^{(\epsilon+nm)\tau-\epsilon\tau^\p}\; D^{ab}(n;p) \,.
\label{Fourier1}
\ee
One recognizes now that
$\epsilon+nm$ is the energy component of the
$4$-momentum $p+nq$ of the incoming particle with the
$O(N)$-index $a$ while
$\epsilon$ is that of $p$ for $b$. In particular, the free propagator in the
mixed representation is
\be
D^{ab}_\om(\tau,\tau^\p)=\delta^{ab}
\int \frac{d\epsilon}{2\pi} \frac{1}{(\epsilon^2-\om^2-i0)}
=\delta^{ab}\frac{1}{2\om} \e^{-\om|\tau-\tau^\p|}
{}~~~~~(\hbox{for }\l=0)\,.
\label{free}
\ee

The equations~\rf{eq2} and \rf{u} can be finally rewritten in
the mixed representation as follows:
\be
\left\{\frac{d^2}{d\tau^2} -\om^2- v(\tau)  \right\}
\frac 1N D^{bb}_\om(\tau,\tau^\p)= -\delta(\tau-\tau^\p)
\label{feq2}
\ee
and
\be
v(\tau) = \l \Phi^2(\tau) +
 \frac{\l}{2\pi^2}\int_{m^2}d\om \;\om \sqrt{\om^2-m^2}
D^{bb}_\om(\tau,\tau) \,.
\label{fu}
\ee

To understand the meaning of the summed amplitude
$D^{bb}_\om(\tau,\tau^\p)$, let us note that
$D^{bc}_\om(\tau,\tau^\p)$ has in the index space the structure
\be
D^{bc}_\om(\tau,\tau^\p) =\left(\delta^{bc}
-\frac{\xi^b\xi^c}{\xi^2}\right)
G^T_\om(\tau,\tau^\p)+ \frac{\xi^b\xi^c}{\xi^2}
G^{S}_\om(\tau,\tau^\p)\,.
\label{structure}
\ee
The amplitudes $G^T$ and $G^{S}$ are associated, respectively, with
the {\it tensor\/} and {\it singlet\/}
$O(N)$-states of two incoming particles.
The averaged over the $O(N)$-indices of two
incoming particles amplitude is
\be
\frac{1}{N} D^{bb}_\om(\tau,\tau^\p) =
\left(1- \frac{1}{N}\right)G^T_\om(\tau,\tau^\p) +
\frac 1N G^{S}_\om(\tau,\tau^\p)
\ee
while the generating function for the symmetrized  over all the $n$+$2$
$O(N)$-indices amplitudes is given by $G^{S}$:
\be
D^{ab}_\om(\tau,\tau^\p) \frac{\xi^a \xi^b}{\xi^2}
= G^{S}_\om(\tau,\tau^\p)\,.
\ee
One sees that  the averaged amplitude which enter  \eq{fu}
coincides with $G^T$  at large $N$.

Substituting \rf{structure} into \eq{feq2}, we rewrite it at large $N$ as
the following equation for $G^T_\om(\tau,\tau^\p)$:
\be
\left\{\frac{d^2}{d\tau^2} -\om^2- v(\tau) \right\}
G^T_\om(\tau,\tau^\p)=
-\delta(\tau-\tau^\p) \,,
\label{eqforG}
\ee
while $v(\tau)$ is related to $G^T_\om(\tau,\tau)$ by
\be
v(\tau) = \l \Phi^2(\tau) +
 \frac{\l N}{2\pi^2}\int_{m^2}d\om \;\om \sqrt{\om^2-m^2} \;
G^T_\om(\tau,\tau) \,.
\label{uvsG}
\ee
Therefore, only the tensor amplitude enters at large $N$.

Eqs.~\rf{eqforG}, \rf{uvsG}
together with \eq{eq1} form the closed set of
equations which will be solved
in the next section.

\newsection{The exact solution}

To solve the set of equations~\rf{eq1}, \rf{eqforG} and \rf{uvsG}, let
us first look at \eq{eqforG} for  given $v(\tau)$. This equation
determines the Green function
of the Schr\"odinger operator with the
potential $v(\tau)$ while $\tau$ plays the role of a $1$-dimensional
coordinate.
In other language $G^T_\om(\tau,\tau^\p)$ is the
matrix element of the resolvent
\be
G^T_\om(\tau,\tau^\p) = \langle \tau |
\frac{1}{-D^2+\om^2+v}|\tau^\p\rangle
\ee
where $D$ stands for $d/d\tau$ for brevity.
One should take then the diagonal matrix element of the resolvent,
$G^T_\om(\tau,\tau)$, in order to substitute into \eq{uvsG}
and to determine $v(\tau)$ versus $\Phi^2$.

The general solution of this problem for  arbitrary $v$
is given by the Gelfand--Diki\u{\i} formula~\cite{GD75}
\be
G^T_\om(\tau,\tau) = R_\om[v] \equiv \sum_{l=0}^\infty
\frac{R_l[v]}{\om^{2l+1}}
\label{GD}
\ee
where the differential polynomials $R_l[v]$ are determined
recurrently by
\be
R_l[v] = \frac{1}{2^l} \left( \frac 12 D^2-v- D^{-1}vD\right)^l
\label{polynomial}
\cdot \frac 12
\ee
and the inverse operator is
\be
D^{-1} v(\tau) = \int^\tau_{-\infty} dx \;v(x) \,.
\ee
\eq{polynomial} stems from the fact that $R_\om[v]$ obeys
the third  order linear differential equation
\be
\fr 12 \left( \fr 12 D^3 -Dv -vD \right)R_\om[v] =
\om^2 DR_\om[v]\,.
\label{linear}
\ee

The polynomials $R_l[v]$ depend on $v$ and its derivatives
$v^{(s)}\equiv (D^s v)$.
The  first few polynomials are
\bea
R_0[v] = \frac 12\;,~~~ R_1[v] = - \frac v4\;,
{}~~~ R_2[v] = \frac {1}{16} (3v^2-v^{\p\p})\;, \non
 R_3[v] = -\frac {1}{64} (10v^3 -10vv^{\p\p} -5(v^\p)^2 +v^{(4)})
\label{RRR}
\eea
while for  $\tau$-independent $v(\tau)=v_0$ one has
\be
R_\om[v_0] = \frac{1}{2\sqrt{\om^2+v_0}}
{}~~~~~\hbox{(} v_0=\hbox{const.)}
\label{constant}
\ee
which agrees with \eq{free} at $\tau=\tau^\p$.

Let us first show how our equations recover the known
results~\cite{Wil73} about the large-$N$ limit of the
$O(N)$-symmetric $\phi^4$ theory.  We put $\xi^a\ra0$
in order to suppress the particle production.  Then the solution
for $v(\tau)$ is $\tau$-independent. Using \eq{constant}
and denoting
\be
v_0 = m^2_R - m^2\,,
\label{mR}
\ee
one rewrites \eq{uvsG} in the form
\be
m^2=m^2_R -
\frac{\l N}{4\pi^2}\int_{m_R^2}d\om \; \sqrt{\om^2-m^2_R} \,.
\label{m2}
\ee
This formula exactly coincides with the standard
expression for the
bare mass, $m$, via the renormalized  mass, $m_R$, at large
$N$~\cite{Wil73}. The meaning of this result is very simple:
since $\xi^a=0$, the only diagrams which are left in the
recurrence relations of Fig.~\ref{fig4} are those
with the  bubble insertions to the propagator. They result
solely in the mass renormalization.

It is convenient to perform the mass renormalization in Eqs.~\rf{eq1},
\rf{eqforG} and \rf{uvsG} eliminating the $\tau$-independent
part of $v$. We introduce
\be
v_R(\tau) = v(\tau) +m^2-m^2_R
\label{uR}
\ee
after which everything is expressed in terms of
$m_R$ and $v_R(\tau)$:
\be
\left\{D^2 - m_R^2 -v_R(\tau) \right\} \Phi^a(\tau) =0\,,
\label{eq1R}
\ee
and
\be
v_R(\tau) = \l \Phi^2(\tau) +
 \frac{\l N}{2\pi^2}\int_{m^2_R}d\om \;\om \sqrt{\om^2-m^2_R}
\left(R_\om[v_R] -\frac{1}{2\om}\right)\,,
\label{uvsGR1}
\ee
where
\be
\om = \sqrt{\vec{p}\;^2+m_R^2} \,.
\ee

We introduce also
the renormalized coupling constant, $\l_R $, which is
related to the bare one, $\l$, by
\be
\frac{1}{\l} = \frac{1}{\l_R}- \frac{N}{8\pi^2}\int_{m_R^2} d\om\,
 \frac {\sqrt{\om^2-m^2_R}} {\om^2}
\label{lR}
\ee
in order to rewrite \eq{uvsGR1} in the form
\be
v_R(\tau) = \l _R\Phi^2(\tau) +
 \frac{\l_R N}{2\pi^2}\int_{m^2_R}d\om \;\om \sqrt{\om^2-m^2_R}
\left(R_\om[v_R] -\frac{R_1[v_R]}{\om^3}-\frac{1}{2\om}\right)\,,
\label{uvsGR}
\ee
The integral over $\om$ on the r.h.s.\ becomes convergent after
the renormalizations.

\eq{lR} coincides with the standard renormalization of the coupling
constant at large $N$~\cite{Wil73}.
The meaning of renormalization is that one chooses the bare
quantities, $m^2$ and $\l$, to be dependent on the cut-off according
to Eqs.~\rf{mR} and \rf{lR} to make the renormalized ones,
$m^2_R$ and $\l_R$, to be cut-off-independent.

I  found the following {\it exact\/} solution to Eqs.~\rf{eq1R}
and~\rf{uvsGR}:
\be
G^T_\om(\tau,\tau)
 = \frac{1}{2\om} -\frac{\bar{\l}_R\Phi^2(\tau)}{4\om(\om^2-m_R^2)} \,,
\label{exact}
\ee
\be
v_R(\tau) =
\bar{\l}_R \Phi^2(\tau)
\label{exact1}
\ee
where
\be
\bar{\l}_R= \frac{\l_R}{1+\frac{\l_R N}{8\pi^2}}
\label{lbar}
\ee
differs from $\l_R$ by a finite factor and
$\Phi^a(\tau)$ reads
\be
\Phi^a(\tau) =
 \frac{\xi^am_R \e^{m_R\tau}}
{1- \frac{\bar{\l}_R \xi^2}{8}\e^{2m_R\tau}} \,.
\label{finalPhi}
\ee

Some comments concerning the exact solution are in order:
\begin{itemize} \vspace{-8pt}
\addtolength{\itemsep}{-8pt}
\item[i)]
Eqs.~\rf{exact}, \rf{exact1} recover
the large-$N$ limit of the results~\cite{Smi93a} for
the tree level $G^T_\om(\tau,\tau)$ and the one-loop level
$\Phi^a(\tau)$.
\item[ii)]
The fact that~\rf{exact}
has a pole only at $\om^2=m_R^2$
means the nullification of  the tensor
on-mass-shell amplitudes $2$$\ra$$n$
for $n$$>$$2$. An analogous property holds~\cite{AKP93b}
at $N=1$ for the tree level amplitudes
in the case of the sine-Gordon potential and for the
$\phi^4$-theory with spontaneously broken symmetry~\cite{Smi93}.
\item[iii)]
The bare coupling $\l$ is related to $\bar{\l}_R$ according to
Eqs.~\rf{lR} and \rf{lbar} as follows
\bea
&(\l N)^{-1} = (\bar{\l}_R N)^{-1}- \frac{1}{8\pi^2}
\left( \int_{m_R^2} d\om\, \om^{-2} \sqrt{\om^2-m^2_R} \;
+1\right) &
\eea
The additional finite renormalization is familiar from the one-loop
result~\cite{Vol93a} for $N=1$.
\item[iiii)]
Since $\bar{\l}_R$ becomes infinite at the negative value
$\l_R=-8\pi^2/N$, $v_R(\tau)$ vanishes at this point
when one goes around the pole singularity in~\rf{finalPhi}.  This is presumably
associated with the double scaling limit of the $4$-dimensional
$O(N)$-symmetric $\phi^4$ theory~\cite{DKO92}.
\item[iiiii)]
The large-$N$
amplitude~\rf{finalPhi} is real which is related to the property of
nullification discussed in the item \/ii).  This is the difference with the
$N=1$ one-loop result~\cite{Vol93a}.
\vspace{-8pt}
\end{itemize}

\begin{figure}[tbp]
\unitlength=1.00mm
\linethickness{0.8pt}
\centering
\begin{picture}(118.00,68.00)(15,70)
\put(23.00,78.00){\makebox(0,0)[cc]{{\Large a)}}}
\put(5.00,128.00){\line(1,0){18.00}}
\put(5.00,100.00){\line(1,0){18.00}}
\put(23.00,100.00){\line(0,1){28.00}}
\put(21.00,114.00){\makebox(0,0)[rc]{{\large $(0,\sqrt{3})$}}}
\put(24.50,128.00){\line(2,1){12.00}}
\put(24.50,100.00){\line(2,-1){12.00}}
\put(24.50,128.00){\line(2,-1){12.00}}
\put(24.50,100.00){\line(2,1){12.00}}
\put(70.00,78.00){\makebox(0,0)[cc]{{\Large b)}}}
\put(70.00,114.00){\line(-2,1){18.00}}
\put(70.00,114.00){\line(-2,-1){18.00}}
\put(71.50,114.00){\line(2,1){18.00}}
\put(71.50,114.00){\line(1,-1){18.00}}
\put(81.50,105.50){\line(4,-1){8.00}}
\put(81.50,105.50){\line(2,1){8.00}}
\put(78.00,105.00){\makebox(0,0)[rc]{{\large $(3,0)$}}}
\put(125.00,114.00){\line(-2,1){18.00}}
\put(125.00,114.00){\line(-2,-1){18.00}}
\put(126.50,114.00){\line(2,-1){18.00}}
\put(126.50,114.00){\line(1,1){18.00}}
\put(133.00,124.00){\makebox(0,0)[rc]{{\large $(3,0)$}}}
\put(125.00,78.00){\makebox(0,0)[cc]{{\Large c)}}}
\put(136.50,122.50){\line(4,1){8.00}}
\put(136.50,122.50){\line(2,-1){8.00}}
\end{picture}
\caption[x]   {\hspace{0.2cm}\parbox[t]{13cm}
{\small
   The tree diagrams for $D^{bb}(4;p)$ at large $N$.
   The components of the $4$-momenta are given for the
   propagating virtual particles
   (in the units of mass) for the on-mass-shell kinematics.
   Only the energy and one space components are written since
   two other components vanish for all particles.
   The momenta of the incoming particles
   are $(2,\sqrt{3})$ and $(2,-\sqrt{3})$, respectively
   while that of the produced
   particles is $(1,0)$. The cancellation of three diagrams
   is illustrated by \eq{cancellation}. }}
   \label{fig5}
   \end{figure}
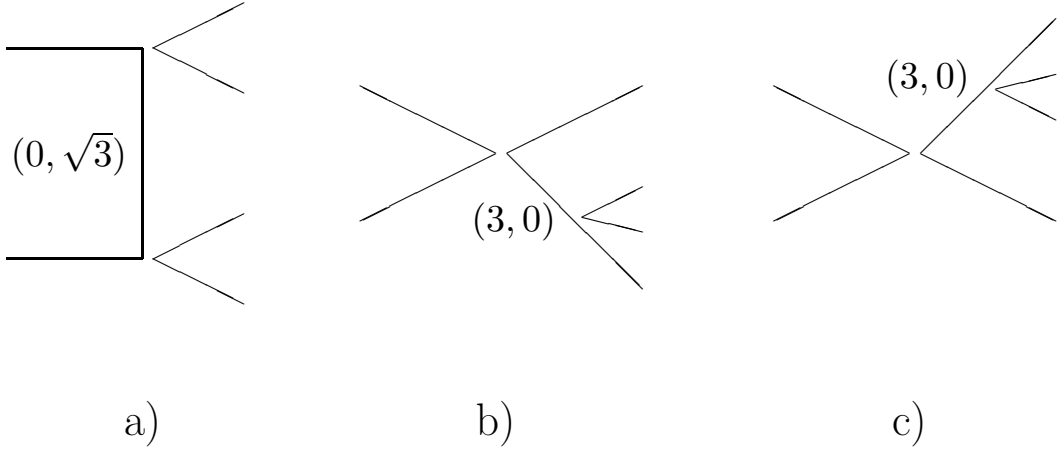

Let us explicitly demonstrate the property listed in
the item \/ii) \/which is crucial for our solution by
the tree diagrams for the process $2$$\ra$$4$.
The diagrams which contribute to $D^{bb}(4;p)$ at large $N$
are depicted in Fig.~\ref{fig5} together with the values
of the $4$-momenta (measured in the units of $m$)
of each propagating virtual particle for the given kinematics
when the incoming particles are on mass shell.
Evaluating the contribution of each diagram of
Fig.~\ref{fig5}$a$, $b$ and $c$
and summing up,
one gets
\be
a) + b) + c) = -\frac 14 +\frac 18 + \frac 18 = 0 \,.
\label{cancellation}
\ee
I performed also an analogous calculation for the
process $2$$\ra$$6$.

Let us now verify  by direct calculations
that~\rf{exact}, \rf{exact1} and \rf{lR} is
 the solution to Eqs.~\rf{uvsGR}, \rf{eq1R}. The proof of the
fact that this solution is unique for the problem of
multiparticle production at threshold is presented in
Appendix~A.

It is instructive  first to verify that $\Phi^2$ is an
eigenvector of the operator
on the r.h.s.\ of \eq{polynomial}:
\be
\fr{1}{2} \left( \fr 12 D^2-v_R- D^{-1}v_RD \right) \Phi^2 =
\fr{1}{2} D^{-1}\left( \fr 12 D^3-Dv_R- v_RD \right) \Phi^2
=D^{-1}Dm_R^2 \Phi^2 =m_R^2 \Phi^2 \,.
\label{eigenvector}
\ee
This formula is derived using solely \eq{eq1R} for an arbitrary
$v_R(\tau)$ and the fact that $\Phi^2(\tau)$ vanishes for
$\tau\ra\-\infty$.

We look now at \eq{exact1} as an ansatz with some
(yet to be determined) coefficient $\bar{\l}_R$.
Then \eq{eigenvector} transforms into
the following  second-order nonlinear differential
equation for $v_R(\tau)$:
\be
v_R^{\p\p} -3v_R^2 = 4m_R^2 v_R \,.
\label{nonlinear}
\ee

Using \eq{RRR} one can rewrite \eq{nonlinear} as
\be
R_2[v_R]=  m_R^2 R_1[v_R] \,.
\label{remarkable}
\ee
This equation is remarkable because
 we can apply again the same operator as in \eq{eigenvector}
to both sides of \eq{remarkable}  to get, according to
\eq{polynomial}, $R_3[v_R]$
on the l.h.s\ and $R_2[v_R]$ on the r.h.s.\ which
is proportional in our case to $v_R$. Applying the operator several
times, we get
\be
R_l[v_R]=-\frac{1}{4}m_R^{2l-2} v_R
\label{powers}
\ee
which gives the formula~\rf{exact} by  using  \eq{GD}.

Substituting \eq{exact} into \eq{uvsGR}, one finds that it is
satisfied for any $\tau$ providing $\l_R$ and $\bar{\l}_R$ are
related by \eq{lbar}.

It is still left to solve \eq{nonlinear} to find an
explicit solution for $v_R(\tau)$. In fact
\eq{nonlinear} is well known and has the solution
which coincides (for $\alpha<0$) with the profile of the one-soliton
solution of the Korteweg-de Vries equation at a fixed value of
time~\cite{New85}:
\be
v_R(\tau) = 2 D^2 \log{\left(1-\alpha \e^{2m_R\tau}\right)}=
\frac{8\alpha m_R^2 \e^{2m_R\tau}}
{\left(1-\alpha \e^{2m_R\tau}\right)^2}
= \frac{2m_R^2}{\hbox{sinh}^2\,(m_R \tau+\fr 12 \log{\alpha})}\,,
\ee
where $\alpha$ is an arbitrary constant which is related to this
fixed value of time and $4m_R^2$ plays the role
of the asymptotic soliton speed.
Since $v_R(\tau)$
is related to $\Phi^2(\tau)$ by \eq{exact1},  one gets
\be
\Phi^a(\tau) =
\sqrt{\frac{8\alpha}{\bar{\l}_R \xi^2}} \frac{\xi^am_R \e^{m_R\tau}}
{\left(1-\alpha \e^{2m_R\tau}\right)} \,.
\label{solforPhi}
\ee

The  expression~\rf{solforPhi} is familiar from the tree level
results~\cite{AKP93a,Bro92}. It satisfies the classical equation
for the $O(N)$-symmetric $\phi^4$ theory with the coupling
constant $\bar{\l}_R$:
\be
\left\{D^2 - m_R^2 -\bar{\l}_R\Phi^2(\tau) \right\} \Phi^a(\tau) =0\,,
\label{eqforPhi}
\ee
which can be easily obtained substituting~\rf{exact1} in
 \eq{eq1R}.

The value of $\alpha$ can be fixed comparing with the
$\xi$$\ra$$0$ limit
when the multiparticle production is suppressed.
In this limit only the diagrams with one insertion of the bubble
chain contribute to the recurrence relation of Fig.~\ref{fig3}.
These diagrams lead, as is discussed above, to the mass
renormalization which is given by \eq{mR}.
In the given case they change the coefficient $m$ in
front of the exponential in the first term on the r.h.s.\ of
\eq{Phi} to be $m_R$. Therefore, the expansion of $\Phi^a$
in $\xi$ should start from
\be
\Phi^a(\tau) =
{\xi^a m_R \e^{m_R\tau}} +  \xi^a {\cal O}(\xi^2)
\label{expansion}
\ee
which fixes $\alpha=\bar{\l}_R \xi^2/8$ and results in \eq{finalPhi}.

Our method of finding the diagonal
resolvent can be extended to the equation
\be
\left\{\frac{d^2}{d\tau^2} -\om^2- 3 v_R(\tau) \right\}
G_\om(\tau,\tau^\p)=
-\delta(\tau-\tau^\p) \,,
\label{eqforGSS}
\ee
where the operator
 differs from the one in \eq{eqforG} by the extra
coefficient $3$ in front of $v_R(\tau)$.
This operator emerges~\cite{Vol93a} in the $N=1$ case.
First we express $G_\om(\tau,\tau)$
via the diagonal resolvent of the operator on the l.h.s.\ of
\eq{eqforGSS}:
\be
G_\om(\tau,\tau) =R_\om[3v_R]
\label{3vG}
\ee
for $v_R$ being the solution of \eq{nonlinear}. We then verify that
\bea
\fr{1}{2} \left( \fr 12 D^2-3v_R- 3D^{-1}v_RD \right) v_R
& = & m_R^2 v_R - \frac 32 v_R^2~, \non
\fr{1}{2} \left( \fr 12 D^2-3v_R- 3D^{-1}v_RD \right) v_R^2
& =& 4m_R^2  v_R^2~.
\label{3v}
\eea
The second equation is remarkable because it shows that  $v_R^2$
is the eigenvector. Applying the operator several times, one proves by
induction that
\be
R_{l+1}[3v_R] = -\frac{3v_R}{4} m_R^{2l} +
(4^l-1)\frac{3v_R^2}{8}m_R^{2l-2}
\ee
which results using \eq{3vG} in the formula
\be
 R_\om[3v_R]= \frac{1}{2\om} -
\frac{3 v_R(\tau)}{4\om(\om^2-m_R^2)} +
\frac{9 v_R^2(\tau)}
{8\om(\om^2-m_R^2)(\om^2-4m_R^2)}\,.
\label{exactSS}
\ee

It is easy to extend the results of the previous paragraph to the
resolvent $R_\om[\frac{s(s+1)}{2}v_R]$ for $s>2$ which also appears in
applications~\cite{Vol93b,BZ93,AKP93d}.  One notices that
\be
\fr{1}{2} \left( \fr 12 D^2-\frac{s(s+1)}{2}v_R-
\frac{s(s+1)}{2}D^{-1}v_RD \right) v_R^l 
 = l^2 m_R^2  v_R^l +
\frac{l(l+1)-s(s+1)}{2(l+1)} \left( l+\fr 12\right) v_R^{l+1}.
\ee
Therefore, the highest power of $v_R$ which emerges in
$R_\om[\frac{s(s+1)}{2}v_R]$ is $v_R^s$ since the second
term on the r.h.s.\ vanishes for $l=s$.

\newsection{Discussion}

An interesting property of the multiparticle production
at threshold in
the $O(N)$-symmetric $\phi^4$ theory at large $N$ is that
the amplitude $a^b(n)$ forms a closed set of exact equations
together with the tensor amplitude $G^T(n;p)$.
The $1$$\ra$$n$ amplitude
$a^b(n)$ is real in a perfect agreement with unitarity
since non-trivial intermediate states are suppressed by
$1/N$ (like the wave-function renormalization starts
from the order $1/N$).
This is similar to the case of $N=1$ with spontaneously broken
symmetry where the $1$$\ra$$n$ amplitude
is real to all orders~\cite{Vol93c}.
As far as $G^T(n;p)$ on mass shell is concern,
it vanishes for $n$$>$$2$ for dynamical reasons
and is real for the process $2$$\ra$$2$ due to kinematics.
I would say that the explicit results for $a^b(n)$
and $G^T(n;p)$
are not very instructive for these reasons for
evaluations of the cross section. Analogously, the
contribution of classical
solutions to the partition function is exponentially
suppressed by $N$ at large $N$.

\begin{figure}[tbp]
\unitlength=1.00mm
\linethickness{0.6pt}
\centering
\begin{picture}(118.00,68.00)(15,70)
\put(23.00,78.00){\makebox(0,0)[cc]{{\Large a)}}}
\put(5.00,128.00){\line(1,0){18.00}}
\put(5.00,100.00){\line(1,0){18.00}}
\put(23.00,101.50){\line(0,1){25.00}}
\put(23.00,128.00){\line(2,1){12.00}}
\put(23.00,100.00){\line(2,-1){12.00}}
\put(23.00,126.50){\line(2,-1){12.00}}
\put(23.00,101.50){\line(2,1){12.00}}
\put(70.00,78.00){\makebox(0,0)[cc]{{\Large b)}}}
\put(70.00,114.00){\line(-2,-1){18.00}}
\put(70.00,115.50){\line(-2,1){18.00}}
\put(70.00,115.50){\line(2,1){18.00}}
\put(70.00,114.00){\line(1,-1){18.00}}
\put(80.00,105.50){\line(4,-1){8.00}}
\put(80.00,105.50){\line(2,1){8.00}}
\put(125.00,119.50){\line(-2,1){18.00}}
\put(125.00,119.50){\line(2,1){18.00}}
\put(125.00,108.50){\line(-2,-1){18.00}}
\put(125.00,108.50){\line(2,-1){18.00}}
\put(125.00,78.00){\makebox(0,0)[cc]{{\Large c)}}}
\put(125.00,114.00){\circle{8.00}}
\put(130.50,114.00){\line(2,1){12.50}}
\put(130.50,114.00){\line(2,-1){12.50}}
\end{picture}
\caption[x]   {\hspace{0.2cm}\parbox[t]{13cm}
{\small
   Some large-$N$ diagrams for $G^{S}(4;p)$. }}
   \label{fig6}
   \end{figure}
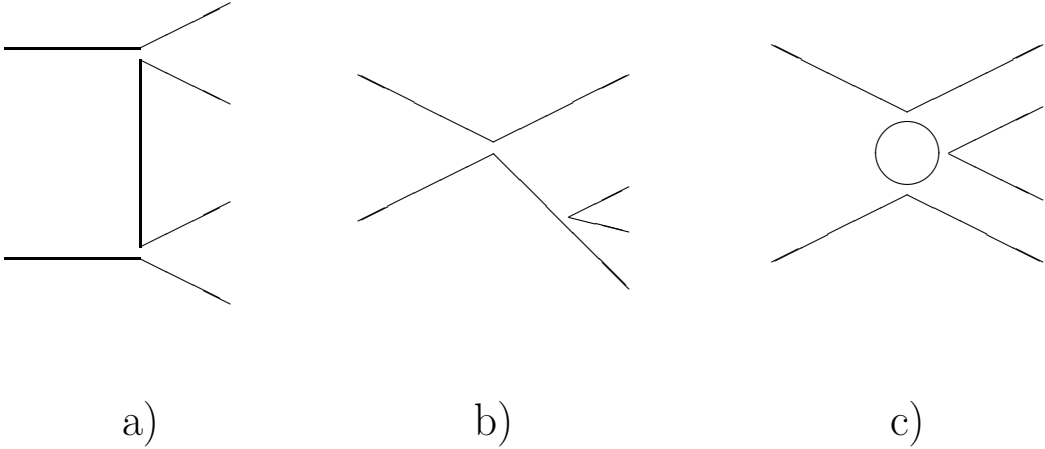

The singlet amplitude $G^S(n;p)$
reveals, on the contrary, a
surprisingly non-trivial behavior even at large $N$.
Some of the diagrams which contribute to $G^S(4;p)$
are depicted in Fig.~\ref{fig6}.
The combinatorics is now
different from that of the diagrams of Fig.~\ref{fig5}
so  there is no cancellation of tree diagrams
for $G^S(4;p)$ on mass shell which happens $n$$\geq$$6$
according to the explicit result~\cite{Smi93a} for $G^S(4;p)$
at the tree level.
Each of vertices can be
``dressed'' with bubble chains and each of the lines
of outgoing particles can be substituted by the exact
amplitude $a^b(n)$ to obtain a diagram with more
produced particles.

One can integrate $G^S(4;p)$ over the $4$-momentum $p$
to obtain diagrams of the vertex type. Some of them are
depicted in Fig.~\ref{fig7}.
\begin{figure}[tbp]
\unitlength=1.00mm
\linethickness{0.6pt}
\centering
\begin{picture}(118.00,72.00)(15,70)
\put(23.00,78.00){\makebox(0,0)[cc]{{\Large a)}}}
\put(23.00,128.00){\line(-5,-4){17.50}}
\put(23.00,100.00){\line(-5,4){17.50}}
\put(23.00,101.50){\line(0,1){25.00}}
\put(23.00,128.00){\line(2,1){12.00}}
\put(23.00,100.00){\line(2,-1){12.00}}
\put(23.00,126.50){\line(2,-1){12.00}}
\put(23.00,101.50){\line(2,1){12.00}}
\put(75.00,78.00){\makebox(0,0)[cc]{{\Large b)}}}
\put(75.00,132.00){\line(-5,-4){22.50}}
\put(75.00,96.00){\line(-5,4){22.50}}
\put(77.00,111.50){\line(0,1){5.00}}
\put(75.00,132.00){\line(2,1){12.00}}
\put(75.00,96.00){\line(2,-1){12.00}}
\put(77.00,111.50){\line(1,0){10.00}}
\put(77.00,116.50){\line(1,0){10.00}}
\put(77.00,129.00){\circle{5.00}}
\put(77.00,120.00){\circle{5.00}}
\put(77.00,99.00){\circle{5.00}}
\put(77.00,108.00){\circle{5.00}}
\put(77.00,104.60){\makebox(0,0)[cc]{$\vdots$}}
\put(77.00,125.50){\makebox(0,0)[cc]{$\vdots$}}
\put(125.00,132.00){\line(-5,-4){22.50}}
\put(125.00,96.00){\line(-5,4){22.50}}
\put(127.00,114.00){\circle{5.00}}
\put(125.00,132.00){\line(2,1){12.00}}
\put(125.00,96.00){\line(2,-1){12.00}}
\put(130.50,114.00){\line(2,1){6.50}}
\put(130.50,114.00){\line(2,-1){6.50}}
\put(127.00,129.00){\circle{5.00}}
\put(127.00,120.00){\circle{5.00}}
\put(127.00,99.00){\circle{5.00}}
\put(127.00,108.00){\circle{5.00}}
\put(127.00,104.60){\makebox(0,0)[cc]{$\vdots$}}
\put(127.00,125.50){\makebox(0,0)[cc]{$\vdots$}}
\put(125.00,78.00){\makebox(0,0)[cc]{{\Large c)}}}
\end{picture}
\caption[x]   {\hspace{0.2cm}\parbox[t]{13cm}
{\small
   The diagrammatic representation of some diagrams for $G^{S}(4;p)$
   at large $N$ integrated over $d^4 p$.
   The diagrams {\normalsize b)} and {\normalsize c)} are of the
   renormalon type
   while {\normalsize $\vdots$} stands for the bubble chain. }}
   \label{fig7}
   \end{figure}
The most interesting are the diagrams
of Fig.~\ref{fig7}$b$,~$c$ which are of the type of
renomalons~\cite{Hoo77} and should behave to $k$-th
order of perturbation theory as $k!$.
It would be very interesting for this reason
to calculate $G^{S}(n;p)$ exactly at large $N$.
This calculation should be simpler than at $N=1$
since the factorization is no longer valid at finite $N$
when one should deal with the whole chain of the
Schwinger--Dyson equations for multipoint Green functions
which were analyzed at the tree level by
Smith~\cite{Smi93b}.

Another interesting model to investigate is the
case of the {\it matrix\/} Higgs field which is described
by the Lagrangian
\be
{\cal L} = \frac 12 \tr{ (\partial_\mu \phi) ^2}
-\frac{m^2}{2}  \tr {\phi ^2}
-\frac{\l_3}{3} \tr {\phi ^3}-
\frac{\l_4}{4}  {\tr \phi ^4}
\label{mlagrangian}
\ee
where $\phi^{ij}(x)$ is generically $N\times N$ matrix.
The model greatly simplifies as $N\ra \infty$ at fixed
$\l_3^2 {N}$ or $\l_4 N$ when only the planar diagrams
survive  similar to  the 't~Hooft limit of QCD~\cite{Hoo74}.
I have checked that the nullification of on-mass-shell
amplitudes holds in the case of the matrix cubic interaction
for $2$$\ra$$3$ at large-$N$. This looks quite similar to what is
observed in this paper for the $O(N)$-case.
The large-$N$ limit of the matrix $\phi^4$ theory
while being simpler than the $N=1$ case
is, however,
quite non-trivial like the large-$N$ limit of QCD  and the
cross section will be no longer suppressed by $1/N$.

\subsection*{Acknowledements}

I am grateful to J.~Ambj{\o}rn, M.~Axenides, P.~Di~Vecchia,
A.~Mironov and P.~Olesen for useful discussions.
This work was sponsored by the Danish Natural Science
Research Council.

\eop

\setcounter{section}{0}
\setcounter{subsection}{0}
\appendix{The proof of uniqueness of the solution}

Let us apply to \eq{uvsGR}  the operator
$\frac{1}{2} \left( \fr 12 D^2-v_R- D^{-1}v_RD \right)$
which enters the recurrence relation~\rf{polynomial}.
The crucial observation is that $\Phi^2$ is an
eigenvector of this operator:
\be
\fr{1}{2} \left( \fr 12 D^2-v_R- D^{-1}v_RD \right) \Phi^2 =
m_R^2 \Phi^2 \,.
\label{egv}
\ee
This formula is derived in \eq{eigenvector}
for an arbitrary $v_R(\tau)$ using only \eq{eq1R}
and the asymptotics of $\Phi^2(\tau)$  for
$\tau\ra\-\infty$.
Therefore, one obtains from \eq{uvsGR}
\be
-4R_2[v_R] = \l_R m^2_R\Phi^2(\tau) +
 \frac{\l_R N}{2\pi^2}\int_{m^2_R}d\om \,\om^3 \sqrt{\om^2-m^2_R}
\left(R_\om[v_R] -\frac{R_2[v_R]}{\om^5}
-\frac{R_1[v_R]}{\om^3}-\frac{1}{2\om}\right).
\label{2}
\ee
Here we used \eq{polynomial} and the fact that $R_\om[v_R]$
satisfies the linear equation~\rf{linear}.

The application of the operator successively $l$ times yields
\be
-4R_{l+1}[v_R] = \l_R m^{2l}_R\Phi^2(\tau) +
 \frac{\l_R N}{2\pi^2}\int_{m^2_R}d\om \;\om^{2l} \sqrt{\om^2-m^2_R}
\left(\om R_\om[v_R] -
\sum_{s=0}^{l+1}\frac{R_s[v_R]}{\om^{2s}}\right) \,.
\label{l}
\ee
Notice that the integral over $\om$ is  convergent
for any $l$.
Multiplying by $\nu^{2l+2}$ and summing over $l$ we get finally
\bea
-4\nu R_{\nu}[v_R] +2 &= &
\frac{\l_R \Phi^2(\tau)}{\nu^2-m_R^2}  \non & &+
 \frac{\l_R N}{2\pi^2}\int_{m^2_R}d\om \;
\frac{\sqrt{\om^2-m^2_R}}{\om^2}\left(
\frac{\om^3 R_{\om}[v_R]-
\nu^3 R_{\nu}[v_R] }{\nu^2-\om^2}  + \frac 12
\right)\,.
\label{fl}
\eea

The last expression
 can be conveniently rewritten as  the contour integral
\be
-4\nu R_{\nu}[v_R] +2 =
\frac{\l_R \Phi^2(\tau)}{\nu^2-m_R^2} +
 \frac{\l_R N}{2\pi^2}\int_{m^2_R}d\om \;
\frac{\sqrt{\om^2-m^2_R}}{\om^2}
\oint_{C_1} \frac{dz}{2\pi i} \frac{z\sqrt{z}
 R_{\sqrt{z}}[v_R] }{(\nu^2-z)(z-\om^2)}
\label{flz}
\ee
where $C_1$ encircles the poles at $z=\om^2$,  $\nu^2$
and  infinity leaving
outside singularities of $\sqrt{z} R_{\sqrt{z}}[v_R]$.

Now the idea is to solve \eq{flz} for  $\nu R_{\nu}[v_R] $ as a function
of $\nu$ rather than  as a functional of $v_R(\tau)$.
The analytic properties of $\sqrt{z}R_{\sqrt{z}}[v_R]$ as a function of $z$
are known for our process of multiparticle production
--- it may have poles and branch cuts on the real axis since
the operator in \eq{eqforG} is self-adjoint. They may start
at $z=(n/2)^2m_R^2$ which is associated with creation of $n$
particles.
In addition, $\sqrt{z}R_{\sqrt{z}}[v_R]$ is complex conjugate
in the upper and lower half-planes since it is real
in the interval $-m_R^2$$<$$z$$<$$m_R^2$ (below any thresholds).
One can always compress the contour $C_1$ in \eq{flz} to encircle
the poles and branch cuts of  $\sqrt{z}R_{\sqrt{z}}[v_R]$ which
yields
\be
-4\nu R_{\nu}[v_R] +2 =
\frac{\l_R \Phi^2(\tau)}{\nu^2-m_R^2} +
 \frac{\l_R N}{2\pi^2}\int_{m^2_R}d\om \;
\frac{\sqrt{\om^2-m^2_R}}{\om^2}
\int \frac{dx}{\pi } \frac{x \im{}\sqrt{x}
 R_{\sqrt{x}}[v_R] }{(\nu^2-x)(x-\om^2)}
\label{flx}
\ee
where the integral is along the support of
$ \im{}\sqrt{z} R_{\sqrt{z}}[v_R] $.

Let us now take the imaginary part of \eq{flx} for real $\nu$. One gets
\be
-4\im \nu R_{\nu}[v_R]  =
{\l_R \Phi^2(\tau)}{\pi\delta(\nu^2-m_R^2)} +
 \frac{\l_R N}{2\pi^2}\int_{m^2_R}d\om \;
\frac{\sqrt{\om^2-m^2_R}}{\om^2}
 \frac{\nu^2 \im{}\nu
 R_{\nu}[v_R] }{(\nu^2-\om^2)} \,.
\label{flim}
\ee
This linear equation for $\im \nu R_{\nu}[v_R]$ has the unique
solution
\be
\im \nu R_{\nu}[v_R] =
{\bar{\l}_R \Phi^2}{\pi\delta(\nu^2-m_R^2)}
\ee
where $\bar{\l}_R$ is given by \eq{lbar}.
Using the requirement that $\nu R_{\nu}[0] =1/2$,
we unambiguously obtain
\be
R_{\nu}[v_R]
 = \frac{1}{2\nu} -\frac{\bar{\l}_R\Phi^2(\tau)}{4\nu(\nu^2-m_R^2)}
\label{ex}
\ee
which coincides with \rf{exact}.

This completes the proof of the uniqueness of the solution~\rf{exact},
\rf{exact1}, \rf{lbar}, \rf{finalPhi}.

\eop

\end{document}